# A Higgs Boson Composed of Gauge Bosons

F. J. Himpsel

Department of Physics, University of Wisconsin Madison,
1150 University Ave., Madison, WI 53706, USA, fhimpsel@wisc.edu

**Abstract**

It is proposed to replace the Higgs boson of the standard model by a Lorentz- and gauge-invariant combination of SU(2) gauge bosons. A pair of Higgs bosons is identified with pairs of gauge bosons by setting their mass Lagrangians equal to each other. That immediately determines the mass of the composite Higgs boson. It becomes simply half of the vacuum expectation value of the standard Higgs boson, which matches the observed mass with tree-level accuracy (2%). The two parameters of the standard Higgs potential are replaced by five one-loop self-interactions of the SU(2) gauge bosons, derived from the fundamental gauge couplings. The Brout-Englert-Higgs mechanism of spontaneous symmetry breaking is generalized from scalars to vectors. Their transverse components acquire finite vacuum expectation values which generate masses for both gauge bosons and the Higgs boson. This concept leads beyond the standard model by enabling calculations of the Higgs mass and its potential without adjustable parameters. It can be applied to non-abelian gauge theories in general, such as grand unified models and supersymmetry.

**Contents**

## 1. A Higgs Boson Composed of Gauge Bosons

The standard model [1] has been highly successful in describing the phenomenology of particle physics. It has passed many high precision tests with flying colors. But the intrinsic elegance of the electroweak gauge theory is blemished by the *ad-hoc* addition of the Higgs field. Rather than letting the gauge symmetry determine all the fundamental bosons, one has to justify the extra Higgs boson empirically. To make the situation worse, a term representing an imaginary mass is introduced into the Lagrangian of the Higgs field, together with a quartic term. Both are unheard of for Lagrangians of fundamental fields. These terms are inserted to obtain an attractive Higgs potential at small field amplitudes and a repulsive potential at large amplitudes. This combination is needed to generate a non-zero vacuum expectation value (VEV).

The discovery of a Higgs-like particle with a mass of about 126 GeV [2] does not alleviate these concerns about an *ad-hoc* Higgs scalar and its artificial potential. A potential escape from this dilemma is the notion of a composite Higgs boson, particularly one that is composed of known particles. In such models the gauge symmetry is broken dynamically by interactions between the constituents of the Higgs boson and in particular by the formation of a condensate similar to the Cooper pairs in superconductivity. A broad class of such models uses a condensate of fermion-antifermion pairs [3],[4]. Since the strength of the interaction between the Higgs boson and fermions is proportional to the fermion mass in the standard model, the heavy top quark has been favored for dynamical symmetry breaking (top quark condensation [4]). These pairing models are able to produce masses for the top quark and the Higgs boson. But the masses come out too large − even after adjusting the inherent high-energy cutoff parameter $\Lambda$. The need for an energy parameter arises from a mismatch in dimensionality between the Higgs boson and a fermion pair. (Bosons have dimension (mass)$^1$ and fermions (mass)$^{3/2}$, in units of $\hbar, c$). Another cause for concern is the short lifetime of the top quark, which prevents the formation of bound states. This problem remains after including the bottom quark to have complete a SU(2) doublet [4].

The model proposed here involves pairing, too, but instead of fermion pairs we consider pairs of gauge bosons. Furthermore, the result of pairing is not an individual Higgs boson, but a pair of Higgs bosons. That guarantees a match of dimensions.

Figure 1 provides more specific heuristics for defining a composite Higgs boson, using diagrams from the standard model. In all three panels a pair of outgoing Higgs bosons on the left side is compared to pairs of outgoing SU(2) gauge bosons on the right. In (a) there are two incoming Higgs bosons, in (b) only one, and in (c) none. Removing the incoming Higgs bosons creates a relation between $H^2$ and $(Z^2+W^+W^-)$ in all three cases. Since the quadratic mass Lagrangians in (c) contain only pairs, they look attractive for defining a pair of composite Higgs bosons from pairs of gauge bosons.

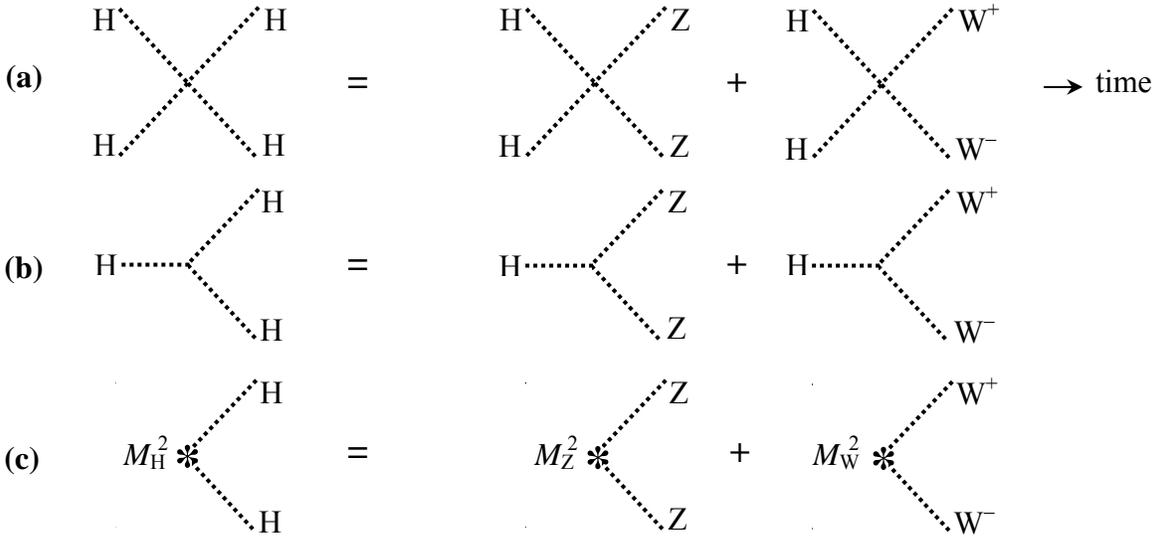

**Figure 1** Standard model diagrams which suggest replacing a pair of outgoing Higgs bosons (in the 1st column) by pairs of outgoing SU(2) gauge bosons (in the 2nd and 3rd columns). The incoming particles are reduced from two in (a) to one in (b) and zero in (c). That suggests a definition of the composite Higgs boson via mass Lagrangians.

The concept of gauge bosons forming a Higgs boson is broadly applicable, since every gauge theory contains them. While they are Lorentz four-vectors, their scalar products match a pair of Higgs scalars. Calculations of gauge boson scattering amplitudes indicate that their interaction is attractive when they form a Lorentz scalar and a singlet of the gauge symmetry (zero spin and isospin) [5]. A further connection between the gauge and Higgs fields is established by the equivalence theorem [6] which connects longitudinal SU(2) gauge bosons with the Goldstone components of the Higgs field. Here we connect transverse SU(2) gauge bosons with the observable Higgs particle. These heuristic considerations lead to a strategy for replacing the Higgs boson of the standard model by a composite of SU(2) gauge bosons:

1) Eliminate the Higgs field from the Lagrangian of the standard model.

2) Define a composite Higgs boson from gauge bosons.

3) Establish a gauge boson potential from one-loop gauge interactions.

4) Obtain symmetry-breaking VEVs and masses by minimizing the potential.

5) Transfer these results from gauge bosons to the composite Higgs boson.

The Higgs field of the standard model can be written as a combination of a SU(2) singlet $H_0$ and a triplet of Goldstone modes $(w_1, w_2, w_3)$, forming a complex doublet:

$$\text{(1)} \quad \Phi_0 = \frac{1}{\sqrt{2}} \begin{bmatrix} w_2 + i w_1 \\ H_0 - i w_3 \end{bmatrix} \qquad \boldsymbol{\Phi}_0 = \mathbf{1} \cdot H_0 + i \Sigma_i \boldsymbol{\tau}^i \cdot w_i \qquad \Phi_0 = \boldsymbol{\Phi}_0 \cdot \begin{bmatrix} 0 \\ 1 \end{bmatrix}$$

$$\Phi_0 = \langle \Phi_0 \rangle + \Phi \qquad \langle \Phi_0 \rangle = \frac{1}{\sqrt{2}} \langle H_0 \rangle \cdot \begin{bmatrix} 0 \\ 1 \end{bmatrix}$$

$$H_0 = \langle H_0 \rangle + H \qquad \langle H_0 \rangle = v = 2^{-1/4} G_F^{-1/2} = 246.2 \text{ GeV}$$

The subscript zero indicates fields with finite VEVs. The complex doublet can be written as 2×2 matrix $\boldsymbol{\Phi}_0$ defined via the Pauli matrices $\boldsymbol{\tau}^i$ and the 2×2 unit matrix $\mathbf{1}$. (All 2×2 matrices will be shown in bold.) The singlet $H_0$ acquires a finite VEV $\langle H_0 \rangle = v$ via the Brout-Englert-Higgs mechanism [7], while the VEVs of the Goldstone modes vanish. The VEV $v$ is directly related to the experimental value of the four-fermion coupling constant $G_F$. After subtracting the VEVs from $\Phi_0, H_0$ one obtains the observable fields $\Phi, H$. The standard Higgs potential combines a quadratic and a quartic term:

$$\text{(2)} \quad V_\Phi = -\mu^2 \cdot \Phi_0^\dagger \Phi_0 + \lambda \cdot (\Phi_0^\dagger \Phi_0)^2 \qquad \text{General gauge}$$

$$V_H = -\tfrac{1}{2} \mu^2 \cdot H_0^2 + \tfrac{1}{4} \lambda \cdot H_0^4 \qquad \text{Unitary gauge}$$

These potentials are reduced from 4$^{th}$ order to 2$^{nd}$ order when using the pairs $\Phi_0^\dagger \Phi_0$ and $H_0^2$ as variables – again a hint that pairs may play a role in Higgs interactions.

The SU(2) gauge bosons form a triplet $(W_\mu^1, W_\mu^2, W_\mu^3)$ that matches the Goldstone triplet. The sum of gauge boson pairs $\Sigma_i W_\mu^i W^{i,\mu}$ is a Lorentz scalar and a SU(2) singlet that matches $\Phi^\dagger \Phi$ up to a proportionality factor p:

$$\text{(3)} \quad \Phi^\dagger \Phi = \tfrac{1}{2} [H^2 + \Sigma_i w_i^2] = -p \cdot \tfrac{1}{2} \Sigma_i W_\mu^i W^{i,\mu}$$

The minus sign ensures that the terms on both sides of the proportionality are positive, taking into account the space-like character of gauge bosons (in the $+---$ metric). The term $\Sigma_i W_\mu^i W^{i,\mu}$ still lacks gauge-invariance, and proportionality constant p has yet to be determined. In the end it will turn out to be unity, as shown in the Appendix. The lack of gauge invariance can be remediated by using chiral electroweak Lagrangians [8]-[14]. These form gauge-invariant building blocks which also incorporate mixing between the SU(2) and U(1) gauge bosons $W_\mu^3$ and $B_\mu$. Even though they have been developed mainly for the heavy Higgs limit $M_H \gg v$ (which is now unrealistic), they allow for a gauge-invariant generalization of $\Sigma_i W_\mu^i W^{i,\mu}$. One starts with a nonlinear representation of the Goldstones $w_i$ by casting them in the form of a SU(2) matrix $\mathbf{U}$:

(4)     $\mathbf{U} = \exp(i \Sigma_i \tau^i \frac{w_i}{v})$

The SU(2)×U(1) gauge bosons are then incorporated by defining the gauge-invariant derivative of the matrix $\mathbf{U}$:

(5)     $D_\mu \mathbf{U} = \partial_\mu \mathbf{U} - ig\mathbf{W}_\mu \mathbf{U} + ig'\mathbf{U}\mathbf{B}_\mu$     $\mathbf{W}_\mu = \Sigma_i \tfrac{1}{2}\tau^i W_\mu^i$     $\mathbf{B}_\mu = \tfrac{1}{2}\tau^3 B_\mu$

Thereby the four gauge bosons $W_\mu^i, B_\mu$ have been converted into the 2×2 matrices $\mathbf{W}_\mu, \mathbf{B}_\mu$. The gauge-invariant derivative $D_\mu$ of the matrix $\mathbf{U}$ defines a four-vector $\mathbf{V}_\mu$ which contains all four gauge bosons and their SU(2)×U(1) couplings $g, g'$:

(6)     $\mathbf{V}_\mu = (D_\mu \mathbf{U})\mathbf{U}^\dagger = i \cdot \tfrac{1}{2}[\Sigma_i \tau^i \cdot (2\partial_\mu \tfrac{w_i}{v} - gW_\mu^i) + \tau^3 \cdot g'B_\mu] = -\mathbf{V}_\mu^\dagger$

In $\mathbf{V}_\mu$ the SU(2) gauge bosons $W_\mu^i$ appear together with the derivatives of the Goldstones $w_i$, showing again their close connection. Replacing $\Sigma_i W_\mu^i W^{i,\mu}$ on the right side of (3) by the trace of $\mathbf{V}_\mu \mathbf{V}^\mu$ establishes a gauge-invariant generalization which includes mixing of the SU(2)×U(1) gauge bosons:

(7)     $\text{tr}[\mathbf{V}_\mu \mathbf{V}^\mu] = -g^2 \cdot [(W_\mu^+ - 2\partial_\mu \tfrac{w_+}{v})(W^{-,\mu} - 2\partial_\mu \tfrac{w_-}{v}) + \tfrac{1}{2}(Z_\mu/c_w - 2\partial_\mu \tfrac{w_3}{v})(Z^\mu/c_w - 2\partial_\mu \tfrac{w_3}{v})]$

  $= -g^2 \cdot [W_\mu^+ W^{-,\mu} + \tfrac{1}{2} Z_\mu Z^\mu / c_w^2]$     for the unitary gauge

  $\to -\tfrac{1}{2} g^2 \cdot \Sigma_i W_\mu^i W^{i,\mu}$     for $g' \to 0$

$W_\mu^\pm, Z_\mu$ and $c_w^2 = \cos^2\theta_w = g^2/(g^2 + g'^2)$, $s_w^2 = 1 - c_w^2$ are defined as usual. Multiplication by $-(\tfrac{1}{2}v)^2$ generates the Lagrangian for the tree-level mass of the gauge bosons:

(8)     $L_M^{ZW} = M_W^2 \cdot (W^+ W^-) + \tfrac{1}{2} M_Z^2 \cdot (ZZ)$     with   $\boxed{M_W = \tfrac{1}{2}gv}$   $\boxed{c_w = M_W/M_Z}$

Scalar products are abbreviated by parentheses from now on. The photon does not appear with the SU(2) gauge bosons, because it is massless. Likewise, one can multiply $\Phi^\dagger \Phi$ on the left side of (3) with the same factor $-(\tfrac{1}{2}v)^2$ to obtain the Lagrangian for a scalar mass $\tfrac{1}{2}v$. In the unitary gauge the Goldstones vanish, and one obtains:

(9)     $L_M^H = -\tfrac{1}{2} M_H^2 \cdot H^2$     with   $\boxed{M_H = \tfrac{1}{2}v}$   $\boxed{g = M_W/M_H}$

The scalar mass is assigned to the tree-level mass of the composite Higgs boson. The resulting value $M_H = \frac{1}{2}v = 2^{-5/4}G_F^{-1/2} = 123\,\text{GeV}$ matches the observed Higgs mass of 126 GeV to about 2%. A comparable agreement exists between the tree-level mass of the W gauge boson $M_W = \frac{1}{2}gv = 78.9\,\text{GeV}$ in (8) and its observed mass of 80.4 GeV. Such an accuracy is typical of the tree-level approximation, which neglects loop corrections of the order $\alpha_w = g^2/4\pi \approx 3\%$. It is reassuring to see the Higgs mass emerging directly from the concept of a Higgs boson composed of gauge bosons.

Next we establish a relation between pairs of gauge bosons and a pair of Higgs bosons by setting the quadratic mass Lagrangians (8) and (9) proportional to each other:

(10) $\boxed{L_M^H = p \cdot L_M^{ZW}}$

After dividing both sides by $-(\frac{1}{2}v)^2$ one arrives at a simple, gauge-invariant relation:

(11) $\boxed{\Phi^\dagger \Phi = p \cdot \text{tr}[\mathbf{V}_\mu \mathbf{V}^\mu]}$ General gauge

(12) $\boxed{\frac{1}{2}H^2 = -p \cdot g^2[(W^+W^-) + \frac{1}{2}(ZZ)/c_w^2]}$ Unitary gauge

A finite VEV for the composite Higgs boson implies finite VEVs for the gauge bosons:

(13) $\boxed{W_0^\pm = \langle W_0^\pm \rangle + W^\pm} \quad \boxed{Z_0 = \langle Z_0 \rangle + Z} \quad (\langle W_0^+ \rangle \langle W_0^- \rangle) = -w^2 \quad (\langle Z_0 \rangle \langle Z_0 \rangle) = -z^2$

Those VEVs have to be transverse to satisfy Lorentz- and gauge-invariance of the vacuum, as discussed in more detail in Section 3, Equation (25). Converting (12) from the observable fields $H, W^\pm, Z$ to the Lagrangian fields $H_0, W_0^\pm, Z_0$ yields the relation:

(14) $\frac{1}{2}(H_0^2 - 2vH_0) = -pg^2\{[(W_0^+W_0^-) + \frac{1}{2}(Z_0Z_0)/c_w^2] - [(\langle W_0^+ \rangle W_0^-) + (W_0^+ \langle W_0^- \rangle) + (\langle Z_0 \rangle Z_0)/c_w^2]\}$

Taking the vacuum expectation value one obtains a relation between VEVs of the Higgs and gauge fields. This relation becomes rather simple when neglecting the zero point oscillations around the VEVs, i.e., $\langle (H_0 - v)^2 \rangle = \langle H_0^2 \rangle - v^2 \ll v^2$ and likewise for $w$ and $z$. Otherwise the VEVs would not be noticeable among the fluctuations. The result is:

(15) $\boxed{v^2 \approx p g^2 (2w^2 + z^2/c_w^2)} \quad$ for $\quad \langle H_0^2 \rangle \approx v^2 \quad \langle (W_0^+ W_0^-) \rangle \approx -w^2 \quad \langle (Z_0 Z_0) \rangle \approx -z^2$

## 2. Dynamical Symmetry Breaking via Gauge Boson Self-Interactions

To test whether gauge bosons can trigger dynamical symmetry breaking, we construct a simple model potential. The gauge-invariant chiral Lagrangian (7) naturally leads to 2nd and 4th order terms analogous to those forming the Higgs potential (2):

(16) $\boxed{V_V = -\mu^2 \cdot \text{tr}[(\mathbf{V}_0\mathbf{V}_0)] + \lambda \cdot (\text{tr}[(\mathbf{V}_0\mathbf{V}_0)])^2}$   General gauge

(17) $\boxed{V_V = g^2\mu^2[(W_0^+W_0^-) + \tfrac{1}{2}(Z_0Z_0)/c_w^2] + g^4\lambda[(W_0^+W_0^-) + \tfrac{1}{2}(Z_0Z_0)/c_w^2]^2}$   Unitary gauge

This is not the standard Higgs potential, though, since the conversion (12) from gauge bosons to the composite Higgs boson applies only to the observable fields $W^\pm, Z, H$. The corresponding conversion (14) of the Lagrangian fields $W_0^\pm, Z_0, H_0$ is more complicated.

The gauge boson model potential (17) is plotted versus the two gauge fields in Figure 2a. Thereby we have used the scalar products $(W_0^+W_0^-)$ and $(Z_0Z_0)$ to define the two field variables $w_0 = [-(W_0^+W_0^-)]^{1/2}$ and $z_0 = [-(Z_0Z_0)]^{1/2}$. Similar pair products appeared already in the definition (12) of the composite Higgs boson. The topography of the model potential is rather peculiar, since the minimum is stretched out over a line. A unique minimum has been generated in Figure 2b by reducing the term $(W_0^+W_0^-) \cdot (Z_0Z_0)$ by a factor $^{10}/_{12}$ and increasing the terms $(W_0^+W_0^-)^2$ and $(Z_0Z_0)^2$ by a factor $^{11}/_{10}$.

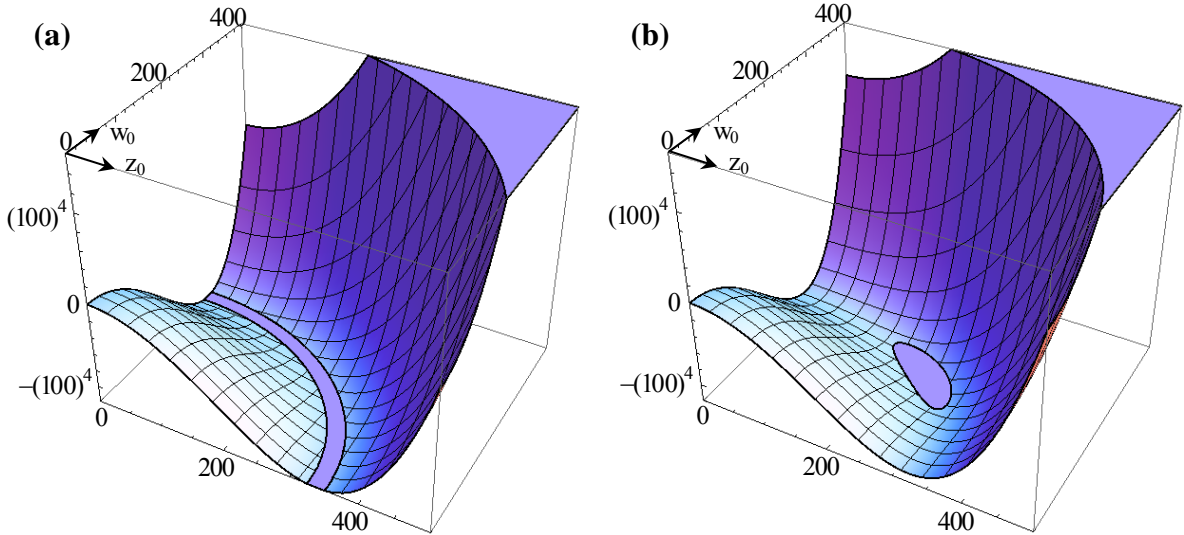

**Figure 2** The potential of the gauge bosons which make up the composite Higgs boson, plotted versus the gauge boson amplitudes $w_0, z_0$. (a) is for the model potential (17) which forms a flat potential valley. (b) is for slightly-modified coefficients which produce a unique minimum. The horizontal axes are in GeV, the vertical axis is in $(\text{GeV})^4$.

Figure 2 demonstrates that scalar products of gauge bosons can develop finite VEVs that break the gauge symmetry. A composite of such scalar products is able to mimic a pair of Higgs scalars. The model potentials still contain the Higgs parameters $\mu^2, \lambda$. The ultimate goal will be a derivation of the gauge boson potential from their quadratic and quartic self-interactions. Those are free of adjustable parameters.

Before tackling this task it is helpful to briefly review the Brout-Englert-Higgs mechanism for converting the Lagrangian Higgs field $H_0$ to the observable Higgs field H. After finding the VEV $v=\mu/\sqrt{\lambda}$ by minimizing the Higgs potential $V_H$ in (2), one changes variables via the substitution $H_0 \rightarrow (v+H)$. The coefficient of $H^2$ provides the mass $M_H$:

(18)  $M_H^2 = 2 \cdot \mu^2$    $v = \mu/\sqrt{\lambda}$

(19)  $V_H = -½\mu^2 \cdot H_0^2 + ¼\lambda \cdot H_0^4$    $H_0 \rightarrow (v+H)$    $v \rightarrow \mu/\sqrt{\lambda}$

   $= -¼\mu^4/\lambda + \mu^2 \cdot H^2 + \mu\sqrt{\lambda} \cdot H^3 + ¼\lambda \cdot H^4$    $\mu \rightarrow M_H/\sqrt{2}$    $\lambda \rightarrow ½M_H^2/v^2$

   $= ½M_H^2 \cdot [-¼v^2 + H^2 + v^{-1} \cdot H^3 + ¼v^{-2} H^4]$

In the last line the parameters $\mu, \lambda$ have been expressed in terms of the observables $M_H, v$. The coefficients $-½\mu^2$ and $¼\lambda$ of the original Higgs potential $V_H(H_0)$ can be retrieved from the quadratic and quartic terms of the converted potential $V_H(H)$. Using the new result $M_H = ½v$ from (9) one can express all the coefficients of the potential by $M_H$:

(20)  $V_H = -½M_H^4 + ½M_H^2 \cdot H^2 + ¼M_H \cdot H^3 + \tfrac{1}{32} \cdot H^4$

The conversion of the Higgs potential from $H_0$ to H is illustrated in Figure 3. The change of the quadratic coefficient from $-½\mu^2 \cdot H_0^2$ to $+½M_H^2 \cdot H^2$ causes the dashed, negative parabola centered at $H_0=0$ to become the dotted, positive parabola at $H=0$. This represents the transition from an attractive potential to a repulsive mass term. A quantitative picture is obtained by inserting experimental values [2] for the observables $M_H = 125.7$ GeV and $v = 2^{-1/4} G_F^{-1/2} = 246.2$ GeV into (18) and solving for $\mu^2, \lambda$:

(21)  $\mu^2 = ½M_H^2 = (88.9\,\text{GeV})^2$    $\lambda = ½M_H^2/v^2 = 0.130$

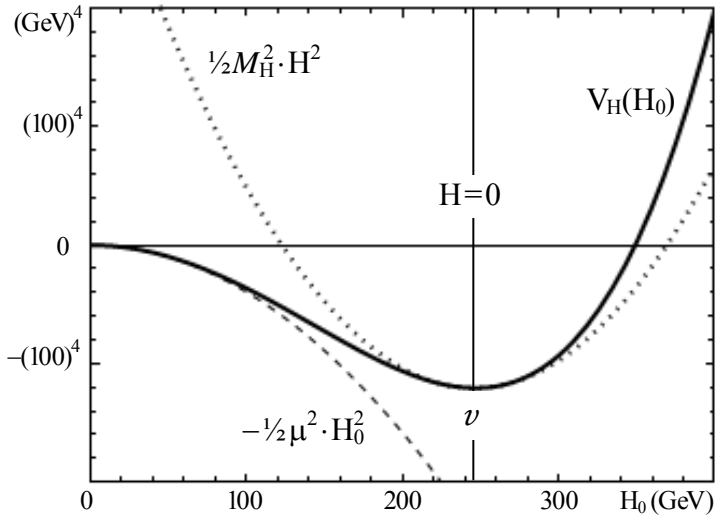

**Figure 3** Plot of the standard Higgs potential $V_H(H_0)$ in (19). The two parameters $\mu^2$ and $\lambda$ are obtained from the observed Higgs mass $M_H$ and the VEV $v$ of $H_0$ via (1),(21). The origin of the observable Higgs boson $H = H_0 - v$ is indicated. $V_H$ has the dimension (mass)$^4$, since it is part of a Lagrangian. In the composite model the dashed potential originates from the gauge boson self-energy.

## 3. Generalization of the Brout-Englert-Higgs Mechanism to Vector Bosons

The gauge boson potential is constructed from one-loop self-interactions containing the neutral pairs ($W_0^+ W_0^-$) or ($Z_0 Z_0$) as external lines. These fields appear in the Lagrangian, like the scalar Higgs field $H_0$ that forms the symmetry-breaking potential (2) of the standard model. Had we chosen the observable gauge bosons $W^\pm, Z$ as variable, we would have encountered odd powers of the fields in the Lagrangian, analogous to the $H^3$ term in the Higgs potential (19).

One-loop diagrams of $O(g^2)$ and $O(g^4)$ are shown in Figure 4 for the standard model and in Figures 5,6 for a pure SU(2) model (which is much easier to handle). Subscripts $_0$ for Lagrangian fields have been suppressed. The 2$^{nd}$- and 4$^{th}$-order terms can be viewed as the first two terms of an infinite series whose N$^{th}$ term is represented by Feynman diagrams of $O(g^{2N})$ containing N pairs of external gauge bosons connected to a loop. The next level of accuracy would involve 6$^{th}$- and 8$^{th}$-order terms, assuming that the trend of alternating signs for even and odd N continues. The Lagrangian fields $W_0^\pm, Z_0$ have large VEVs, around which the observable fields $W^\pm, Z$ oscillate with small amplitudes. For applying perturbation theory there are two options to be considered:

1) Use a truncated perturbation series in the weak observable fields $W^\pm, Z$. There are additional diagrams for odd powers of the fields, such as the triple gauge boson couplings $W^+W^-Z$ and $ZZZ$. It possible to avoid such vertices by retrieving the coefficients of $W_0^\pm, Z_0$ from quadratic and quartic diagrams for $W^\pm, Z$ (compare the discussions after (19) and (26) for the Higgs and gauge bosons potentials).

2) Sum the one-loop series for the strong Lagrangian fields $W_0^\pm, Z_0$ to infinity. That has been accomplished for the quartic Higgs self-interaction [15]. It looks like a formidable task for the gauge bosons of the standard model, where the number of diagrams escalates rapidly with the number of external lines. But it might be feasible for pure SU(2) fields.

The following figures show irreducible diagrams for the observable gauge bosons of the standard model (in the unitary gauge). A complete set of diagrams for a general gauge would have to include reducible diagrams, crossed diagrams, Goldstone modes, gauge fixing terms, Fadeev-Popov ghosts, and counterterms for renormalization.

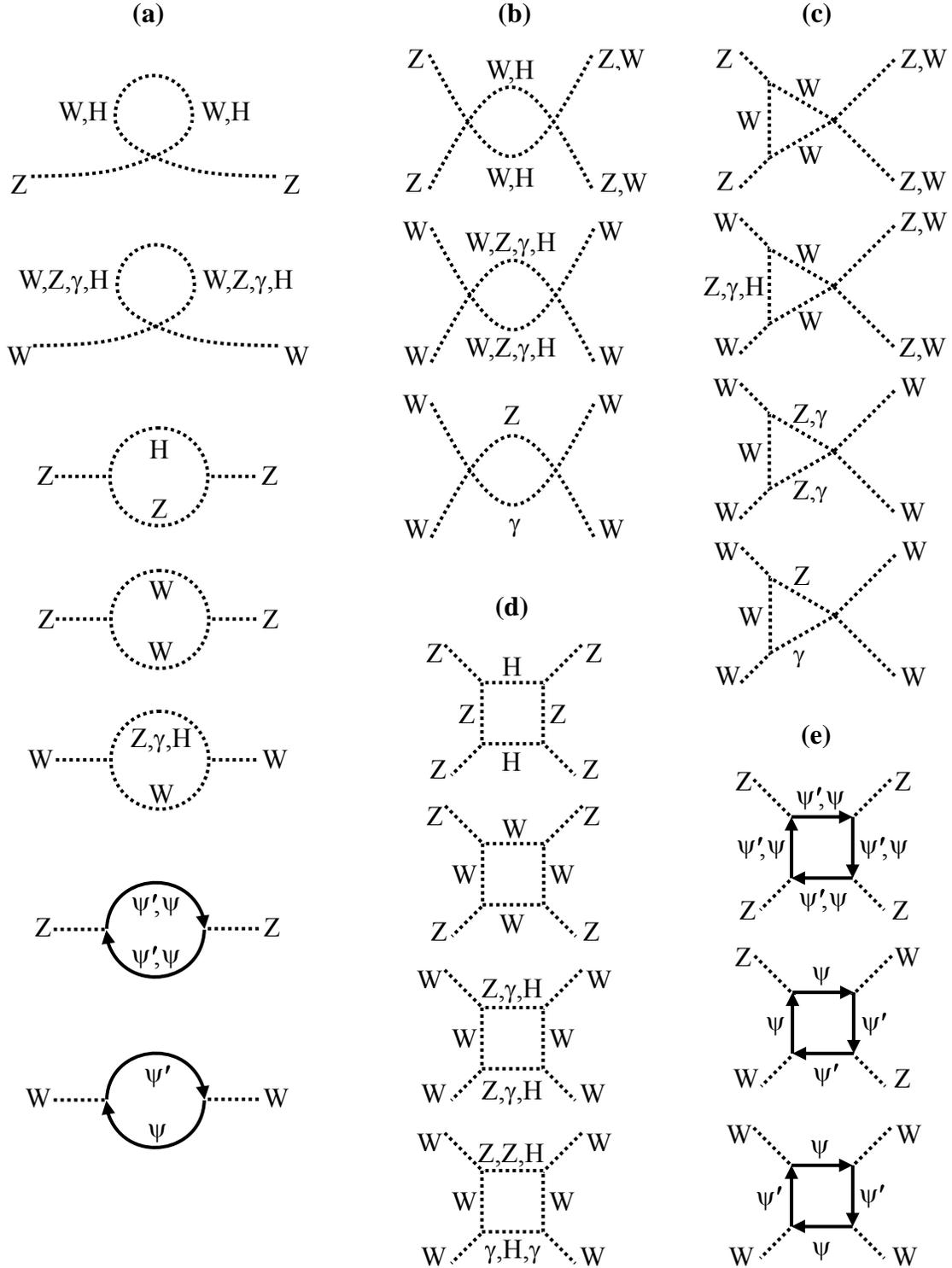

**Figure 4** Irreducible one-loop self-interactions of the observable SU(2) gauge bosons in the standard model: (a) Quadratic diagrams of $O(g^2)$ for the self-energies $\Sigma^W$ and $\Sigma^Z$, (b)-(e) diagrams of $O(g^4)$ representing scattering between the neutral gauge boson pairs $(W_0^+ W_0^-)$ and $(Z_0 Z_0)$. They generate the quadruple vertex corrections $\Lambda^{WW}, \Lambda^{WZ}, \Lambda^{ZZ}$. Left-handed fermion doublets are labeled $(\psi, \psi')$.

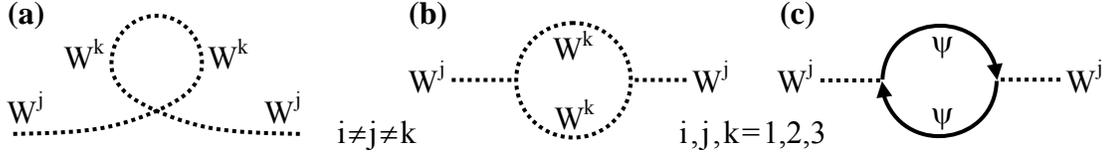

**Figure 5** Quadratic self-interactions of $O(g^2)$ for the pure SU(2) gauge bosons $W^i$: (a) gauge boson loop with a quadruple vertex, (b) gauge boson loop with two triple vertices, (c) fermion loop. Together they represent the gauge boson self-energy $\Sigma$. Such diagrams form the attractive part of the potential.

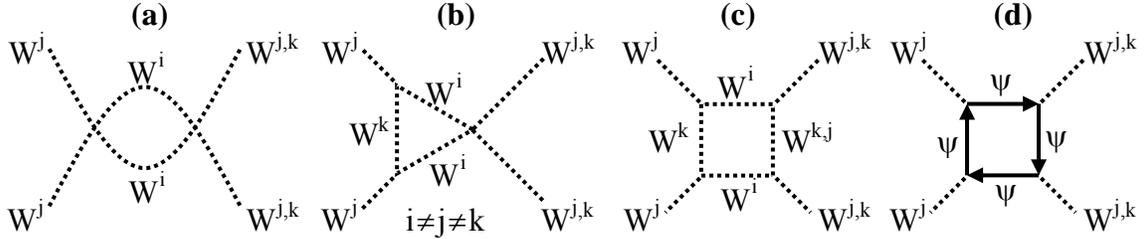

**Figure 6** Quadruple self-interactions of $O(g^4)$ between pure SU(2) gauge bosons, forming the repulsive part of the potential: (a) two quadruple vertices, (b) one quadruple vertex plus two triple vertices, (c) four triple vertices, (d) four vertices with fermions. These diagrams describe the vertex correction $\Lambda$ for scattering between pairs of equal gauge bosons. $\psi$ represents a fermion.

From the combinations of external lines in Figure 4 one obtains the general structure of the dynamic gauge boson potential:

(22)  $V_V^{dyn} = V_{2V} + V_{4V}$

$$V_{2V} = -\Sigma^W \cdot (W_0^+ W_0^-) - \tfrac{1}{2}\Sigma^Z \cdot (Z_0 Z_0)$$

$$V_{4V} = \Lambda^{WW} \cdot (W_0^+ W_0^-)^2 + \tfrac{1}{4}\Lambda^{ZZ} \cdot (Z_0 Z_0)^2 + \Lambda^{WZ} \cdot (W_0^+ W_0^-)(Z_0 Z_0)$$

The factor ½ with $(Z_0 Z_0)$ has been included to ensure that $\Sigma^Z$ adds to $M_Z^2$ in the mass Lagrangian (8). It does not occur with $(W_0^+ W_0^-)$, which represents the two real fields $W^1, W^2$. The same prefactors were kept in defining $V_{4V}$. The model potential (17) corresponds to the coefficients:

(23)  $\Sigma^W = c_w^2 \cdot \Sigma^Z = -\mu^2 g^2$    $\Lambda^{WW} = c_w^2 \cdot \Lambda^{WZ} = c_w^4 \cdot \Lambda^{ZZ} = \lambda g^4$

With $\mu^2, \lambda$ obtained from $M_W, v$ via (21) one arrives at the following coefficients:

$\Sigma^W = -(57.0\,\text{GeV})^2$   $\Sigma^Z = -(64.6\,\text{GeV})^2$   $\Lambda^{WW} = 0.0219$   $\Lambda^{ZZ} = 0.0363$   $\Lambda^{WZ} = 0.0282$

The characteristics of the gauge boson potential come out more clearly by plotting Figure 2 versus the pair variables $w_0^2 = -(W_0^+ W_0^-)$ and $z_0^2 = -(Z_0 Z_0)$, as done in Figure 7. These variables are matched to the scalar products of gauge bosons occurring in the gauge boson potential (22). 4$^{th}$ order terms now become quadratic and 2$^{nd}$ order terms linear. Since the potential has been reduced to a quadratic form, it can be analyzed in terms of quadric surfaces in the three-dimensional space spanned by the variables $x = z_0^2$, $y = w_0^2$, and $z = V_V^{dyn}$. The potential surfaces have paraboloidal character, since the variable z appears only linearly, not quadratically. For dynamical symmetry breaking one needs a potential surface with a minimum, which has the general form of an elliptic paraboloid as shown in Fig. 7b. For the model potential in Section 2 the paraboloid degenerates to a parabolic cylinder (Fig. 7a). The topology of these potential surfaces is determined by the determinant of the coefficient matrix for the 2$^{nd}$ order terms in (22).

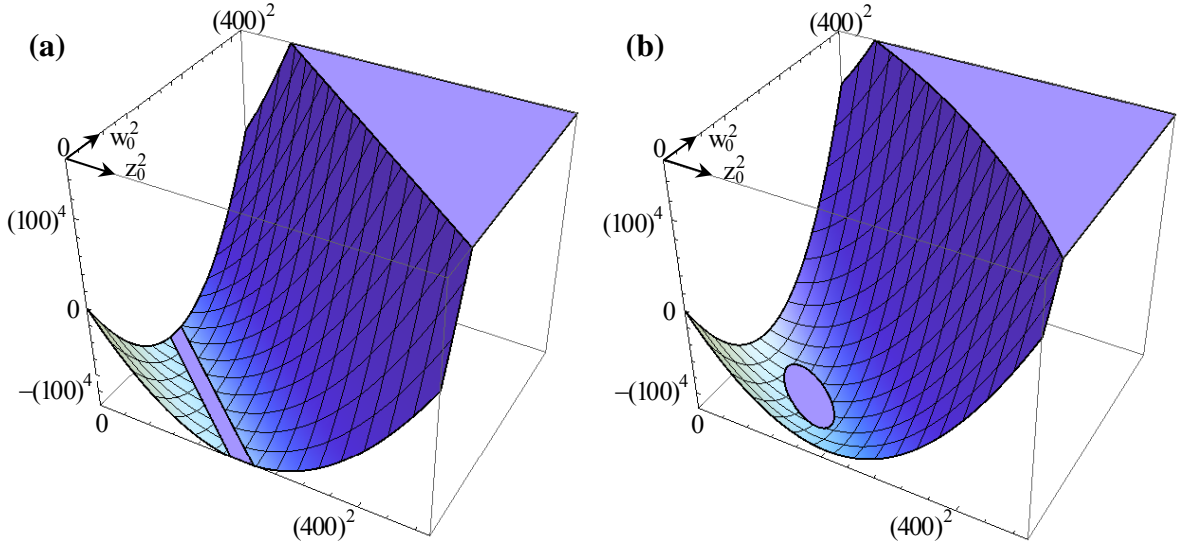

**Figure 7**  Plots of the two gauge boson potentials from Fig. 2, using the squared amplitudes $w_0^2, z_0^2$ as variables instead of $w_0, z_0$. That leads to a simpler topography of the potential. In (a) the extended minimum of the model potential (17),(23) becomes a straight line at the bottom of a parabolic cylinder when plotted against $w_0^2, z_0^2$. For the modified potential in (b) the potential surface becomes an elliptical paraboloid. This simplicity matches the concept of gauge boson pairs as natural variables of the symmetry-breaking potential. The horizontal axes are now in (GeV)$^2$.

The next step consists of minimizing the gauge boson potential with respect to the (positive) variables $-(W_0^+ W_0^-) = w_0^2$ and $-(Z_0 Z_0) = z_0^2$. If a well-defined minimum exists at finite VEVs $w_0^2 = w^2$ and $z_0^2 = z^2$, one can shift $W_0^\pm, Z_0$ to the observable fields $W^\pm, Z$. To be attractive for small amplitudes and repulsive for large amplitudes, the potential needs negative quadratic coefficients $\Sigma^W, \Sigma^Z$ and positive quartic coefficients $\Lambda^{WW}, \Lambda^{ZZ}$ (analogous to $-\mu^2$ and $\lambda$ for the scalar Higgs potential). The quadratic term $V_{2V}$ contains a product of three minus signs. One is explicit and the other two are implicit, because $\Sigma^W, \Sigma^Z$ and the scalar products of vector bosons are negative.

At the minimum of $V_V^{dyn}(w_0^2, z_0^2)$ the partial derivatives with respect to the pair coordinates $w_0^2, z_0^2$ have to vanish. To obtain the proper topology of the potential surface, one has to consider the determinant $\frac{1}{4}[\Lambda^{WW} \cdot \Lambda^{ZZ} - (\Lambda^{WZ})^2]$ of the 2×2 matrix containing the 2$^{nd}$ order coefficients $\Lambda^{WW}$, $\frac{1}{4}\Lambda^{ZZ}$, and $\frac{1}{2}\Lambda^{WZ}$. The degenerate case with vanishing determinant will be postponed to Section 4. For a non-vanishing, positive determinant one obtains a single minimum at the VEVs $w^2, z^2$ under the following conditions:

(24a) $\partial V_V^{dyn}/\partial(w_0^2) = 0 \qquad \partial V_V^{dyn}/\partial(z_0^2) = 0$

(24b) $\Sigma^W, \Sigma^Z < 0 \qquad \Lambda^{WW}, \Lambda^{ZZ} > 0$

(24c) $\Lambda^{WW} \cdot |\Sigma^Z| > \Lambda^{WZ} \cdot |\Sigma^W| \qquad \Lambda^{ZZ} \cdot |\Sigma^W| > \Lambda^{WZ} \cdot |\Sigma^Z| \qquad \Lambda^{WW} \cdot \Lambda^{ZZ} > (\Lambda^{WZ})^2$

(24d) $\boxed{\begin{aligned} w^2 &= -(\langle W_0^+\rangle\langle W_0^-\rangle) = \tfrac{1}{2}[\Lambda^{ZZ}\cdot|\Sigma^W| - \Lambda^{WZ}\cdot|\Sigma^Z|]/[\Lambda^{WW}\cdot\Lambda^{ZZ} - (\Lambda^{WZ})^2] \\ z^2 &= -(\langle Z_0\rangle\langle Z_0\rangle) = [\Lambda^{WW}\cdot|\Sigma^Z| - \Lambda^{WZ}\cdot|\Sigma^W|]/[\Lambda^{WW}\cdot\Lambda^{ZZ} - (\Lambda^{WZ})^2] \end{aligned}}$

When assigning a VEV to a vector boson, one has to be careful to preserve the Lorentz invariance of the vacuum. It would be violated by choosing a fixed vector in space-time. But it is possible to escape this dilemma by assigning the VEV to the longitudinal or transverse components of a vector boson. Thereby the VEV of each individual particle is oriented relative to its momentum vector. Since the infinite number of virtual particles in the vacuum of quantum field theory covers the full range of momenta, individual orientation effects are averaged out globally. But they exist locally within a Compton wavelength around each gauge boson. Since gauge transformations are local, gauge symmetry can be broken locally.

At a first glance, one might be tempted to associate the VEV with the longitudinal component of a vector boson. It exists only when the gauge boson becomes massive after the gauge symmetry is broken by a finite VEV. But such an assignment would be gauge-dependent, since the longitudinal component could be eliminated by the Landau gauge. Its role would be transferred to a Goldstone mode, whose VEV vanishes. As a result, the VEV of the gauge boson would be gauged away. Consequently, the VEV of a gauge boson must be associated with its two transverse components.

Choosing the momentum of a gauge boson as z-axis of a local reference frame, one can convert $W_0^\pm, Z_0$ into scalars, multiplied by one of the two transverse polarization vectors $\varepsilon_{T,n}^\mu$. These scalars are identified with the field amplitudes $w_0, z_0$ that were used already for plotting the potentials. In the Landau gauge one obtains:

(25)
$$\boxed{\begin{array}{l} W_0^\pm = \langle W_0^\pm \rangle + W^\pm \quad\quad \langle W_0^\pm \rangle = w^\pm \cdot \varepsilon_{T,n} \quad\quad W^\pm = w^\pm \cdot \varepsilon_{T,n} \\ Z_0 = \langle Z_0 \rangle + Z \quad\quad\quad \langle Z_0 \rangle = z \cdot \varepsilon_{T,n} \quad\quad\quad Z = z \cdot \varepsilon_{T,n} \end{array}}$$

$$\begin{array}{lll} W_0^1 = \langle W_0^1 \rangle + W^1 & \langle W_0^1 \rangle = w \cdot \varepsilon_{T,n} & W^1 = w^1 \cdot \varepsilon_{T,n} \\ W_0^2 = \langle W_0^2 \rangle + W^2 & \langle W_0^2 \rangle = w \cdot \varepsilon_{T,n} & W^2 = w^2 \cdot \varepsilon_{T,n} \\ w^+ = (w^1 - i w^2)/\sqrt{2} & w^- = (w^1 + i w^2)/\sqrt{2} & w^2 = w^+ w^- = \tfrac{1}{2}[(w^1)^2 + (w^2)^2] \\ w^+ = (w - i w)/\sqrt{2} & w^- = (w + i w)/\sqrt{2} & w^2 = w^+ w^- \\ \varepsilon_{T,1}^\mu = (0,1,0,0) & \varepsilon_{T,2}^\mu = (0,0,1,0) & \varepsilon_{T,m}^\mu \varepsilon_{T,n,\mu} = -\delta_{mn} \quad\quad \varepsilon_T^\mu \cdot k_\mu = 0 \end{array}$$

$$\boxed{-(Z_0 Z_0) = z_0^2 = z^2 + 2 z z + z^2} \quad\quad \boxed{-(W_0^+ W_0^-) = w_0^2 = w^2 + (w^1 + w^2) w + w^2}$$

The VEV of the photon remains zero, since the U(1) symmetry of QED is not broken.

With the structure of the VEVs in hand, one can convert the Lagrangian fields $W_0^\pm, Z_0$ to the observable fields $W^\pm, Z$ in the potential (22). The required substitutions (13) are worked out in the last line of (25). Products of fields and VEVs generate odd powers of the fields, as for the cubic term of the scalar Higgs potential in (19):

(26)
$$\boxed{\begin{array}{l} V_V^{dyn} = \Lambda^{WW} \cdot (w^+ w^-)^2 + \tfrac{1}{4} \Lambda^{ZZ} \cdot (zz)^2 + \Lambda^{WZ} \cdot (w^+ w^-)(zz) \\ + 2 w \Lambda^{WW} (w^1 + w^2)(w^+ w^-) + z \Lambda^{ZZ} \cdot z(zz) + 2 z \Lambda^{WZ} \cdot z(w^+ w^-) + w \Lambda^{WZ} \cdot (w^1 + w^2)(zz) \\ \quad + 2 w^2 \Lambda^{WW} (w^+ w^-) + z^2 \Lambda^{ZZ} \cdot (zz) + 2 w^2 \Lambda^{WW} \cdot (w^1 w^2) + 2 w z \Lambda^{WZ} \cdot (w^1 + w^2) z \\ \quad\quad + \tfrac{1}{2} w^2 \Sigma^W + \tfrac{1}{4} z^2 \Sigma^Z \end{array}}$$

The three scalar fields $w^\pm, z$ describe the transverse components of the observable gauge bosons, as laid out in (25). Since the Feynman diagrams describing (26) have neutral external lines, it is advantageous to choose the neutral gauge bosons $w^1 = (w^+ + w^-)/\sqrt{2}$, $w^2 = i(w^+ - w^-)/\sqrt{2}$, and $z$ as basis set.

Equation (26) is valid in the Landau gauge, where the longitudinal components have been converted to Goldstone modes. To include those in the definition of the composite Higgs boson one has to use the general form (11), which includes Goldstones via (3),(7). If one eliminates the Goldstones by choosing the unitary gauge, the longitudinal gauge bosons and their self-interactions have to be added to (26).

The potential of the observable gauge bosons in (26) resembles the Higgs boson potential in the 2$^{nd}$ and 3$^{rd}$ lines of (19). The constant term corresponds to a change of the zero point energy, which does not matter here. The linear terms vanish, since the origin of the shifted fields lies at the minimum of the potential. The quadratic terms $(w^+w^-),(zz)$ take the form of a vector boson mass Lagrangian (8) after inserting the definitions $w^2 = -(W^+W^-)$, $z^2 = -(ZZ)$:

(27) $\boxed{M_W^2 = 2\Lambda^{WW} \cdot w^2} = \Lambda^{WW} \cdot [\Lambda^{ZZ} \cdot |\Sigma^W| - \Lambda^{WZ} \cdot |\Sigma^Z|] / [\Lambda^{WW} \cdot \Lambda^{ZZ} - (\Lambda^{WZ})^2]$

$\boxed{M_Z^2 = 2\Lambda^{ZZ} \cdot z^2} = 2\Lambda^{ZZ} \cdot [\Lambda^{WW} \cdot |\Sigma^Z| - \Lambda^{WZ} \cdot |\Sigma^W|] / [\Lambda^{WW} \cdot \Lambda^{ZZ} - (\Lambda^{WZ})^2]$

$c_W = M_W/M_Z = (\Lambda^{WW}/\Lambda^{ZZ})^{1/2} \cdot w/z$ (in the on-shell renormalization scheme)

These mass terms are analogs of the result $M_H^2 = 2\lambda \cdot v^2$ for the standard Higgs boson. They are indeed positive with the constraints $\Lambda^{WW}, \Lambda^{ZZ} > 0$ from (24b). The VEVs $w^2, z^2$ can be expressed by the coefficients of the gauge boson potential via (24d), producing lengthy expressions in (27). These show how the self-interactions $\Sigma^W, \Sigma^Z$ and $\Lambda^{WW}, \Lambda^{WZ}, \Lambda^{ZZ}$ of the Lagrangian fields can be extracted from quadratic and quartic terms of the observable fields, as done in the analysis of (19) for the Higgs boson. The quartic coefficients $\Lambda^{WW}, \Lambda^{WZ}, \Lambda^{ZZ}$ do not change during the conversion from Lagrangian to observable fields. They can be obtained from diagrams for observable gauge bosons such as those in Fig. 4 for the standard model. The result is then inserted into the lengthy mass terms of (27), which can be calculated from quadratic diagrams for observable gauge bosons. That determines the quadratic coefficients $\Sigma^W, \Sigma^Z$ of the Lagrangian gauge bosons without having to deal with strong Lagrangian fields that cannot be handled by perturbation theory. The remaining terms in (26) are self-interactions of the gauge bosons analogous to the cubic and quartic Higgs self-interactions.

A test of this model requires calculations of the gauge boson self-interactions. Those are shown in Fig. 4 for the standard model. The quadratic self-energy diagrams have been calculated in various places [16],[17]. But renormalized results are difficult to find in explicit form. A plot of the renormalized transverse self-energies by Böhm et al. 1986 [16] shows that the gauge boson self-energies are negative, as required for an attractive potential that induces spontaneous symmetry-breaking. But the mass of the top quark was highly underestimated at that time, making the results only qualitative.

Another interesting finding in this work is a nearly quadratic increase of the gauge boson self-energies when going away from the mass shell towards higher energies. That gives rise to substantial values when evaluating $\Sigma^W, \Sigma^Z$ at the VEVs $w, z$ around which $W_0^\pm, Z_0$ oscillate. An empirical determination of $\Sigma^W, \Sigma^Z$ and $w, z$ confirms that (see (33) in Section 4). Thus, one-loop radiative corrections can give rise to substantial masses despite their suppression by a factor $g^2$.

Calculations of the quadruple self-interactions $\Lambda^{WW}, \Lambda^{ZZ}, \Lambda^{WZ}$ are still unavailable, even though many results have been published for chiral electroweak Lagrangians in the heavy Higgs limit [8]-[13]. These are not applicable to the light Higgs particle discovered in 2012 [2]. Particluarly interesting would be the Lagrangian $L_5 = \alpha_5 \cdot \mathrm{tr}[\mathbf{V}_\mu \mathbf{V}^\mu]^2$ which describes the quartic term in the model potential (16). More recent approaches should be applicable [14]. Furthermore, one might be able to use studies of the invariant amplitudes for boson-boson scattering. Their next-to-leading terms of $O(g^4)$ are related to the coefficients $\Lambda^{WW}, \Lambda^{ZZ}, \Lambda^{WZ}$. Most calculations have been performed in the high energy limit, spurred by the unitarity catastrophe looming at the TeV energy scale for a heavy Higgs boson [6]. In that limit the longitudinal components of the gauge bosons dominate and the transverse components can be ignored. Nevertheless, exact amplitudes have been calculated for WW→WW and ZZ→ZZ scattering at the one-loop level, involving hundreds of diagrams [18]. It would be interesting to see these amplitudes translated into the coefficients $\Lambda^{WW}$ and $\Lambda^{ZZ}$ for the gauge boson potential.

For further guidance one could consult long-standing efforts to generate the quadratic coefficient $\mu^2$ of the standard Higgs potential from its quartic self-interaction [15],[19],[20]. Calculations carried out within the standard model via the renormalization group equations have found expressions for the Higgs self-energy $\Sigma^H$ of the form:

(28) $\quad \Sigma^H(\Lambda) \;=\; 3(\Lambda/4\pi v)^2 \cdot [M_H^2(\Lambda) + 2M_W^2(\Lambda) + M_Z^2(\Lambda) - 4m_t^2(\Lambda)] \;<\; 0$

For energy scales $\Lambda$ ranging from $v$ all the way up to about $10^{17}$ GeV the result is indeed negative [20], as required for spontaneous symmetry breaking. The sign is dominated by the contribution from the mass $m_t$ of the top quark. It is interesting to notice that fermion loops also dominate the negative self-energies of the gauge bosons [16],[17]. In that case the light fermions dominate.

## 4. Phenomenology

In the absence of calculations for the five coefficients of the gauge boson potential one can at least test whether they can be matched to five observables. Natural candidates are $M_W, M_Z, M_H, v,$ and $\tan\theta_H$ (the mixing ratio between $(W^+W^-)$ and $(ZZ)$ in the definition (12) of the composite Higgs boson). The Higgs VEV $v$ can be obtained directly from the four-fermion coupling $G_F$ via (1). While the previous section was aimed at calculating observables from the coefficients, the goal is now to determine the coefficients from observables. For this purpose we characterize the potential surfaces that can be obtained with various parameter sets. Contour plots, such as those in Figure 8, are a good way to analyze the situation.

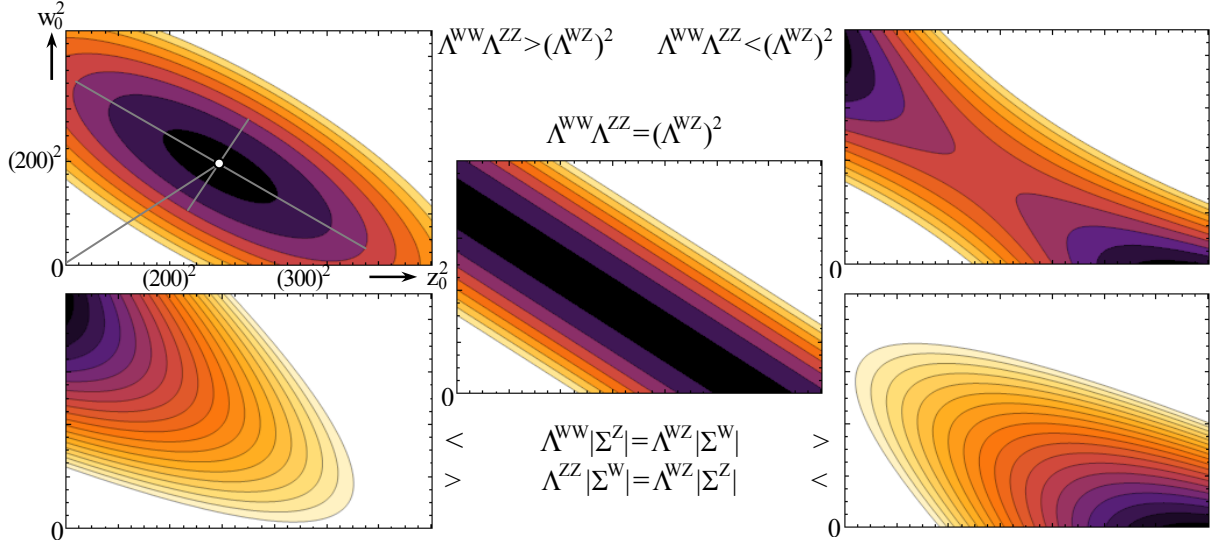

**Figure 8**  The contours of the gauge boson potential (22) as a function of the pair coordinates $w_0^2, z_0^2$ [in $(\text{GeV})^2$]. Minima are shown dark. The panels illustrate various scenarios in terms of the three inequalities (24c). The central panel is for the degenerate model potential (17),(23) with $\Lambda^{WW} \cdot \Lambda^{ZZ} = (\Lambda^{WZ})^2$, where the minimum expands into a line. The upper left panel for $\Lambda^{WW} \cdot \Lambda^{ZZ} > (\Lambda^{WZ})^2$ exhibits a unique minimum with finite VEVs. The opposite inequality $\Lambda^{WW} \cdot \Lambda^{ZZ} < (\Lambda^{WZ})^2$ leads to a saddle point (upper right). That creates a bistable situation with two minima at the edges of the allowed region $(w_0^2, z_0^2 \geq 0)$. When one of the other inequalities in (24c) is violated, a single minimum tends to occur at one of the two edges, i.e., one of the VEVs vanishes (bottom panels).

In general, one can use the three inequalities in (24c) as guidelines for the shape of the contours. They affect the signs of the numerators and denominators in (24d) which in turn determine the gauge boson VEVs $w^2, z^2$. If the inequalities are satisfied, one obtains a well-defined minimum in the $w_0^2, z_0^2$ plane (top left panel). If one has equalities instead, the minimum becomes stretched out into a line, leading to a degenerate potential (central panel). With a negative numerator one of the VEVs gets pushed to the boundary of the allowed quadrant and vanishes (bottom panels). If the common denominator $[\Lambda^{WW}\Lambda^{ZZ}-(\Lambda^{WZ})^2]$ in (24d) turns negative, the minimum becomes a saddle point and drives both VEVs towards the boundaries (upper right panel).

The degenerate case shown at the center of Figure 8 is easier to analyze, because the number of coefficients is reduced from five to three. The relations (24a-d) for the non-degenerate case are replaced by the following set:

(29a) $\partial V_V^{dyn}/\partial(w_0^2)=0 \qquad \partial V_V^{dyn}/\partial(z_0^2)=0$

(29b) $\Sigma^W, \Sigma^Z < 0 \qquad \Lambda^{WW}, \Lambda^{ZZ} > 0$

(29c) $\boxed{\Lambda^{WW}\cdot\Sigma^Z = \Lambda^{WZ}\cdot\Sigma^W \qquad \Lambda^{ZZ}\cdot\Sigma^W = \Lambda^{WZ}\cdot\Sigma^Z} \quad \Rightarrow \quad (\Lambda^{WZ})^2 = \Lambda^{WW}\cdot\Lambda^{ZZ}$

(29d) $\boxed{z^2 = -[\Sigma^W/\Lambda^{WZ} + 2\Sigma^W/\Sigma^Z \cdot w^2]} \qquad \Rightarrow \quad d(w^2)/d(z^2) = -\Sigma^Z/2\Sigma^W$

The two independent constraints in (29c) can be used to eliminate $\Lambda^{WW}, \Lambda^{ZZ}$:

(30) $\boxed{\Lambda^{WW} = \Lambda^{WZ}\cdot(\Sigma^W/\Sigma^Z)} \qquad \boxed{\Lambda^{ZZ} = \Lambda^{WZ}\cdot(\Sigma^Z/\Sigma^W)}$

Equation (29d) describes a line of possible VEVs $w^2, z^2$ in the $w_0^2, z_0^2$ plane (labeled VEV in Figure 9). This introduces an extra degree of freedom, chosen to be $w$. Altogether there are four independent parameters $(\Sigma^W, \Sigma^Z, \Lambda^{WZ}, w)$. These will be narrowed down by experimental constraints in the following.

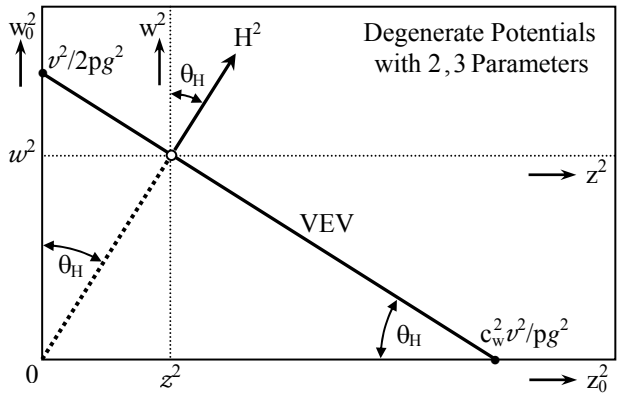

**Figure 9** Analysis of the degenerate potentials (29a-d). The full line covers possible locations of the VEV in the $w_0^2, z_0^2$ plane, with $w_0^2=-(W_0^+W_0^-)$ and $z_0^2=-(Z_0Z_0)$. The arrow for $H^2$ shows that a Higgs pair is a mixture of the gauge boson pairs $w^2=-(W^+W^-)$ and $z^2=-(ZZ)$, with the mixing angle fixed by (12) to $\theta_H=32.8^0$.

Figure 9 contains two independent parts. The upper right quadrant is for the observable fields $w^2, z^2$ and the remainder for the Lagrangian fields $w_0^2, z_0^2$. The two parts are connected at the VEV $w^2, z^2$ (small circle), the origin of the arrow for $H^2$. At this point the two potentials for the Lagrangian and observable fields can be matched (including their derivatives, such as the slope of the VEV line).

From the connection (12) between the observable Higgs and gauge boson fields one learns that the Higgs pair $H^2$ is a linear combination of the gauge boson pairs $w^2 = -(W^+W^-)$ and $z^2 = -(ZZ)$. The rotation angle $\theta_H$ relative to the $w^2$ axis is given by:

(31) $\boxed{\tan\theta_H = 1/2c_w^2 = 0.6435}$ $\quad \theta_H = 32.8^0$

$\sin\theta_H = 1/(1+4c_w^4)^{1/2}$ $\quad \cos\theta_H = 2c_w^2/(1+4c_w^4)^{1/2}$

$\theta_H$ can be viewed as a mixing angle analogous to the electroweak mixing angle $\theta_w$. But it mixes gauge boson pairs instead of individuals. $\theta_H$ and $\theta_w$ are indeed related via $c_w^2 = \cos^2\theta_w$. Since $H^2$ is a mass eigenstate, its arrow has to follow the steepest potential gradient, which is orthogonal to the VEV line. Consequently the slope of the VEV line given in (29d) has to be equal to $\tan\theta_H$ in (31) up to a sign:

(32) $\boxed{\Sigma^Z = \Sigma^W/c_w^2}$ $\qquad c_w^2 = M_W^2/M_Z^2 = 0.777$

That eliminates the parameter $\Sigma^Z$. This constraint yields the model potential (17) in Section 2. The coefficients $\Sigma^W, \Sigma^Z, \Lambda^{WZ}$ can then be mapped onto the two parameters $\mu^2, \lambda$ of the standard Higgs potential via (23).

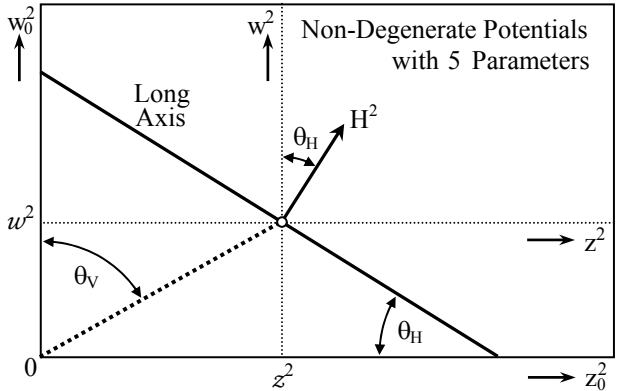

**Figure 10** Similar to Fig. 9, but for the case of non-degenerate potentials. The dashed line pointing towards the VEV $w^2, z^2$ is generally not parallel to the arrow for mass eigenstate $H^2$, making $\theta_V$ different from $\theta_H$. A second massive state appears parallel to the long axis of the elliptical potential contours (compare Fig. 8, top left).

Non-degenerate potentials are shown in the upper left panel of Fig. 8 and analyzed in Fig. 10. Instead of a line of possible VEVs one has now a well-defined point, surrounded by elliptical equipotential contours when plotted in the $w^2, z^2$ plane. Since the composite Higgs boson is a mass eigenstate, the arrow for $H^2$ has to be aligned with one of the two principal axes of the ellipses (gray crosshairs in Fig. 8, full lines in Fig. 10). $\theta_H$ does not coincide anymore with the angle $\theta_V$, as it did in Figure 9.

A possible complication is the appearance of a massive scalar field along the second principal axis of the potential contours. This eigenstate combines WW and ZZ pairs with opposite signs, suggesting an iso-tensor. In that case one expects a repulsive pair interaction [5], which would prevent pair formation. The second eigenstate is not a problem in for degenerate potentials, where it becomes massless. In that case it does not contribute to the definition (10) of the composite Higgs boson via mass Lagrangians.

The Appendix investigates in detail how the five independent coefficients $\Sigma^W, \Sigma^Z$, $\Lambda^{WW}, \Lambda^{ZZ}, \Lambda^{WZ}$ can be matched to the five observables $M_W, M_Z, M_H, v, \tan\theta_H$. Degenerate and non-degenerate potentials behave quite differently. As characteristic result we list the unique set of coefficients and VEVs that can be derived from degenerate potentials:

(33) $\boxed{\begin{array}{ll} \Sigma^W = -(98.45\,\text{GeV})^2 & \Sigma^Z = -(111.7\,\text{GeV})^2 \\ \Lambda^{WW} = 0.0657 \quad \Lambda^{ZZ} = 0.1088 \quad \Lambda^{WZ} = 0.0845 \end{array}}$ $\boxed{w = 222\,\text{GeV} \quad z = 195\,\text{GeV}}$

This set should be taken as an initial estimate. A major uncertainty is the contribution from longitudinal gauge bosons. Their VEVs vanish, but they have their own self-energies [16],[17] and vertex corrections. Although the longitudinal gauge bosons disappear in the Landau gauge, they reappear as Goldstone modes in the Higgs sector according to the equivalence theorem. The Goldstones then enter the gauge-invariant definition (11) of the composite Higgs boson on both sides via (3) and (7).

The gauge boson self-energies $\Sigma^W, \Sigma^Z$ are comparable to $-M_W^2, -M_Z^2$. For the standard Higgs boson the situation is similar. Writing the attractive part of the Higgs potential in Fig. 3 as ½$\Sigma_H \cdot H^2$ one obtains $\Sigma_H = -2\mu^2 = -M_H^2$. Likewise, the quadruple vertex corrections in (33) are comparable to the quartic coupling $\lambda = 0.130$ of the standard Higgs boson in (21). And the gauge boson VEVs are comparable to the Higgs boson VEV.

## 5. Summary and Outlook

In summary, a new concept is proposed for electroweak symmetry breaking, where the Higgs boson is identified with a scalar combination of gauge bosons in gauge-invariant fashion. That explains the mass of the Higgs boson with 2% accuracy. In order to replace the standard Higgs scalar, the Brout-Englert-Higgs mechanism of symmetry breaking is generalized from scalars to vectors. The *ad-hoc* Higgs potential of the standard model is replaced by self-interactions of the SU(2) gauge bosons, which can be calculated without adjustable parameters. Dynamical symmery breaking then leads to finite VEVs of the transverse gauge bosons and thereby generates masses and self-interactions. Gauge bosons and their interactions are related directly to the symmetry group of a theory via the adjoint representation and gauge-invariant derivatives. Therefore the proposed mechanism of dynamical symmetry breaking is applicable to any non-abelian gauge theory, including grand unified theories and supersymmetry.

In order to test this model, five gauge boson self-interactions need to be worked out. These are the self-energies $\Sigma^W, \Sigma^Z$ and the four-fold vertex corrections $\Lambda^{WW}, \Lambda^{ZZ}, \Lambda^{WZ}$. The VEV of the standard Higgs boson – which generates masses for the gauge bosons and for the Higgs itself – gets replaced by the VEVs of the $W^\pm$ and Z gauge bosons. Since the standard Higgs boson interacts with most of the fundamental particles, its replacement implies rewriting a large portion of the standard model. As a first step one could calculate the five self-interactions within the standard model and compare them to the values obtained empirically by matching five observables. The upcoming high-energy run of the LHC offers a great opportunity to test the characteristic couplings of the composite Higgs boson, as well as the new gauge boson couplings introduced by their VEVs. If confirmed, the concept of a Higgs boson composed of gauge bosons would open the door to escape the confine of the standard model and calculate previously inaccessible quantities, such as the Higgs mass and the Higgs potential.

**Appendix:   Obtain Gauge Boson Self-Interactions from Observables**

The gauge boson potential (22) contains five coefficients, two for the self-energies and three for the quadruple vertex corrections. Rather than undertaking the difficult task of calculating them, one can take a phenomenological approach and match them to five observables. There are indeed five natural observables characterizing the SU(2) gauge bosons and the Higgs boson. These fall into two distinct groups. $M_W, M_Z, \tan\theta_H$ depend only on the gauge boson potential, while $M_H, v$ involve the definition of the composite Higgs boson in (12). That brings in the additional proportionality constant p. In the following we will carry out this approach, first for non-degenerate potentials and then for their degenerate limit.

The starting point for non-degenerate potentials is (26), combined with the VEVs in (24d) and the gauge boson masses in (27). The self-energies $\Sigma^W, \Sigma^Z$ can be connected to the masses $M_W, M_Z$ by inserting (24d) into (27) and solving for $\Sigma^W, \Sigma^Z$:

(A1) $\quad \Sigma^W = -(\tfrac{1}{2}M_Z^2 \cdot \Lambda^{WW} + M_W^2 \cdot \Lambda^{WZ})/\Lambda^{WW}$

$\quad\quad\quad \Sigma^Z = -(\tfrac{1}{2}M_Z^2 \cdot \Lambda^{WZ} + M_W^2 \cdot \Lambda^{ZZ})/\Lambda^{ZZ}$

$\Lambda^{WZ}$ is obtained by requiring that one axis of the elliptical contours in Fig. 8 (upper left) is parallel to the arrow for $H^2$ in Fig. 10. The two axes are parallel to the eigenvectors of the symmetric 2×2 matrix with the diagonal elements $\Lambda^{WW}$, $\tfrac{1}{4}\Lambda^{ZZ}$ and the off-diagonal $\tfrac{1}{2}\Lambda^{WZ}$. Multiplying this matrix from both sides with the vector $\{w_0^2 = -(W_0^+ W_0^-),\ z_0^2 = -(Z_0 Z_0)\}$ produces the quartic gauge boson potential $V_{4V}$ in (22) which determines the orientation of the elliptical potential contours in Fig. 8. The eigenvector with two positive signs points in the direction of the $H^2$ arrow. Therefore, its slope in the $w_0^2, z_0^2$ plane is $\tan\theta_H = 1/2c_w^2$ according to (31). The resulting constraint is then solved for $\Lambda^{WZ}$:

(A2) $\quad \{(\Lambda^{WW} - \tfrac{1}{4}\Lambda^{ZZ}) + [(\Lambda^{WW} - \tfrac{1}{4}\Lambda^{ZZ})^2 + (\Lambda^{WZ})^2]^{1/2}\}/\Lambda^{WZ} = 2c_w^2 \quad\quad c_w^2 = M_W^2/M_Z^2$

$\quad\quad\quad \Lambda^{WZ} = (\Lambda^{WW} - \tfrac{1}{4}\Lambda^{ZZ}) \cdot c_w^2/(c_w^4 - \tfrac{1}{4})$

Combining (A1),(A2) produces three relations involving only gauge boson observables:

(A3) $\quad \boxed{\begin{array}{l} \Sigma^W = -M_W^2 \cdot (1 + \tfrac{1}{2}R \cdot M_Z^4/\Lambda^{ZZ}) \\ \Sigma^Z = -M_Z^2 \cdot (\tfrac{1}{2} + R \cdot M_W^4/\Lambda^{WW}) \\ \Lambda^{WZ} = R \cdot M_W^2 M_Z^2 \end{array}} \quad\quad \boxed{R = (\Lambda^{WW} - \tfrac{1}{4}\Lambda^{ZZ})/(M_W^4 - \tfrac{1}{4}M_Z^4)}$

The two remaining constraints are the definition (12) of the composite Higgs boson and the relation (15) for its VEV $v$. The latter yields a relation between $\Lambda^{WW}$ and $\Lambda^{ZZ}$ by solving (27) for the gauge boson VEVs $w^2, z^2$ and inserting those into (15):

(A4) $\quad \boxed{\Lambda^{ZZ} = \tfrac{1}{2}pg^2 \cdot M_Z^4/M_W^2 \cdot \Lambda^{WW}/(v^2 \cdot \Lambda^{WW} - pg^2 \cdot M_W^2)}$

For incorporating the Higgs mass $M_H$ via (12) one has to extract the potential of the composite Higgs boson from the potential of of the gauge bosons. Both potentials involve the observable fields, which makes them more complicated than those of the Lagrangian fields (see (22) and (26) for gauge bosons, and the 1st and 2nd lines in (19) for the standard Higgs boson). For the gauge boson potential (26) one actually needs three independent variables instead of the two variables $\{w^2, z^2\}$ used in the figures. It is convenient to choose the three real gauge fields $(w^1)^2 = -(W^1 W^1)$, $(w^2)^2 = -(W^2 W^2)$, and $z^2 = -(ZZ)$, as defined in (25). The redundant variable $w^2 = (w^+ w^-) = \frac{1}{2}[(w^1)^2 + (w^2)^2]$ needs to be eliminated from (26) in favor of $(w^1)^2, (w^2)^2$. According to the definition (12), the potential of the composite Higgs boson covers a line in the space spanned by $\{(w^1)^2, (w^2)^2, z^2\}$. These variables become functions of $H^2$ after combining (12) with the constraint (31) for $\tan\theta_H$ and the equivalence of $w^1$ and $w^2$:

(A5)  $H^2 = pg^2 \cdot [(w^1)^2 + (w^2)^2 + z^2/c_w^2]$

$1/2c_w^2 = \tan\theta_H = z^2/(w^+ w^-) = 2z^2/[(w^1)^2 + (w^2)^2]$    $(w^1)^2 = (w^2)^2$

Solving for the three coordinates gives:

(A6)  $\boxed{(w^1)^2 = C \cdot H^2 \quad (w^2)^2 = C \cdot H^2 \quad z^2 = \tfrac{1}{2} C \cdot H^2}$    $\boxed{C = 2M_W^4/(M_Z^4 + 4M_W^4) pg^2}$

The gauge boson mass term in (26) is now converted to the Higgs mass term in (19) by the substitution $(w^+ w^-) \to \tfrac{1}{2}[(w^1)^2 + (w^2)^2]$, followed by the substitutions (A6):

(A7)  $V_V^M = \tfrac{1}{2} M_W^2 \cdot (w^1)^2 + \tfrac{1}{2} M_W^2 \cdot (w^2)^2 + \tfrac{1}{2} M_Z^2 \cdot z^2$

$\qquad = \tfrac{1}{2} M_W^2 / pg^2 \cdot H^2$

$V_V^M = \tfrac{1}{2} M_H^2 \cdot H^2$    $\boxed{M_W^2 / pg^2 = M_H^2}$

Setting the two versions of the Higgs mass term equal to each other produces a condition for the proportionality constant p. For the natural value p=1 this becomes exactly the tree-level mass formula for the composite Higgs boson in (9). Inserting the experimental masses [2] one obtains p=0.995. In summary, four out of five coefficients of the gauge boson potential plus the parameter p are fixed by matching the five observables $M_W, M_Z, M_H, v$, and $\tan\theta_H$. The remaining degree of freedom can be assigned to $\Lambda^{WW}$. Its upper limit $\Lambda^{WW} = 0.0657$ corresponds to the degenerate solution.

Degenerate gauge boson potentials were introduced in (29 a-d). Compared to the non-degenerate potentials in (22),(24 a-d) they are constrained by two conditions for $\Lambda^{WW}, \Lambda^{ZZ}$ in (29c), but exibit an extra degree of freedom for $w$ or $z$ in (29d). $\Lambda^{WW}, \Lambda^{ZZ}$, and $\Lambda^{WZ}$ are constrained by (29c), leaving $\Sigma^W, \Sigma^Z, w, p$ as independent variables:

(A8)  $\boxed{\Lambda^{WW} = \Lambda^{WZ} \cdot (\Sigma^W / \Sigma^Z)}$    $\times$    $\boxed{\Lambda^{ZZ} = \Lambda^{WZ} \cdot (\Sigma^Z / \Sigma^W)}$    $\Rightarrow$    $\boxed{(\Lambda^{WZ})^2 = \Lambda^{WW} \cdot \Lambda^{ZZ}}$

The potential (26) for the observable gauge bosons then becomes:

(A9)
$$\begin{aligned}
V_V^{dyn} = &+ \Sigma^W/\Sigma^Z \cdot \Lambda^{WZ} \cdot (w^+w^-)^2 && + \tfrac{1}{4}\Sigma^Z/\Sigma^W \cdot \Lambda^{WZ} \cdot (zz)^2 \\
&+ \Lambda^{WZ} \cdot (w^+w^-)(zz) \\
&+ 2w \cdot \Sigma^W/\Sigma^Z \cdot \Lambda^{WZ} \cdot (w^1+w^2)(w^+w^-) && + z \cdot \Sigma^Z/\Sigma^W \cdot \Lambda^{WZ} \cdot z(zz) \\
&+ 2z \cdot \Lambda^{WZ} \cdot z(w^+w^-) && + w \cdot \Lambda^{WZ} \cdot (w^1+w^2)(zz) \\
&+ 2w^2 \cdot \Sigma^W/\Sigma^Z \cdot \Lambda^{WZ} \cdot (w^+w^-) && + z^2 \cdot \Sigma^Z/\Sigma^W \cdot \Lambda^{WZ} \cdot (zz) \\
&+ 2w^2 \cdot \Sigma^W/\Sigma^Z \cdot \Lambda^{WZ} \cdot (w^1w^2) && + 2wz \cdot \Lambda^{WZ} \cdot (w^1+w^2)z \\
&- \tfrac{1}{4}\Sigma^W\Sigma^Z/\Lambda^{WZ}
\end{aligned}$$

$$z^2 = -\Sigma^W/\Lambda^{WZ} - 2\Sigma^W/\Sigma^Z \cdot w^2 \qquad w^2 = -\tfrac{1}{2}\cdot\Sigma^Z/\Lambda^{WZ} - \tfrac{1}{2}\Sigma^Z/\Sigma^W \cdot w^2$$

The gauge boson masses are extracted from the coefficients of $(w^+w^-),(zz)$:

(A10) $M_W^2 = 2\Sigma^W/\Sigma^Z \cdot \Lambda^{WZ} \cdot w^2$

$M_Z^2 = 2\Sigma^Z/\Sigma^W \cdot \Lambda^{WZ} \cdot z^2$

Now we can match $\Sigma^W, \Sigma^Z, w$ to the observables $M_W, M_Z, \tan\theta_H$ via (A10),(31),(32):

(A11) $\boxed{\Sigma^W = -\tfrac{3}{2} M_W^2} \qquad \boxed{\Sigma^Z = -\tfrac{3}{2} M_Z^2} \qquad \boxed{w^2 = \tfrac{1}{2} M_Z^2/\Lambda^{WZ}}$

Next one has to establish a connection with the composite Higgs boson. Its VEV $v$ is brought in via (15), which connects the parameter p with $\Lambda^{WZ}$:

(A12) $p = \tfrac{2}{3} v^2 \Lambda^{WZ}/g^2 M_Z^2$

The last remaining observable to be matched is the Higgs mass. Thereby we proceed as described in the previous paragraph by evaluating the gauge boson mass term along the Higgs line in the three-dimensional space spanned by the variables $(w^1)^2, (w^2)^2, z^2$. That yields again the requirement $M_W^2/pg^2 = M_H^2$ via (A7), together with the choice p=1. Inserted into (A12) that pins down the last undetermined coefficient $\Lambda^{WZ}$:

(A13) $\boxed{\Lambda^{WZ} = \tfrac{3}{2} g^2 M_Z^2/v^2}$

The complete set of coefficients obtained from the experimental values of the five observables [2] is listed in (33), together with the gauge boson VEVs.